\documentclass[twocolumn,showpacs,prl,amsmath,amssymb]{revtex4}
\usepackage{graphicx}
\usepackage{dcolumn}
\usepackage{bm}

\begin{document}
\title{Phonon anharmonicities in graphite and graphene}

\author{Nicola Bonini$^1$, Michele Lazzeri$^2$, Nicola Marzari$^1$, and Francesco Mauri$^2$}
\affiliation{
$^1$ Department of Materials Science and Engineering, Massachusetts Institute
 of Technology, Cambridge MA 02139, USA\\
$^2$ IMPMC, Universit\'es Paris 6 et 7, CNRS, IPGP, 140 rue de Lourmel, 75015 Paris, France}
\date{\today}

\begin{abstract}
We determine from first-principles the finite-temperature
properties---linewidths, line shifts, and lifetimes---of the key
vibrational modes that dominate inelastic losses in graphitic
materials.  In graphite, the phonon linewidth of the Raman-active
E$_{2g}$ mode is found to decrease with temperature; such anomalous
behavior is driven entirely by electron-phonon interactions, and does
not appear in the nearly-degenerate infrared-active E$_{1u}$ mode.  In
graphene, the phonon anharmonic lifetimes and decay channels of the
A$'_1$ mode at {\bf K} dominate over E$_{2g}$ at ${\bm \Gamma}$ and
couple strongly with acoustic phonons, highlighting how ballistic
transport in carbon-based interconnects requires careful engineering
of phonon decays and thermalization.
\end{abstract}

\maketitle

Carbon nanotubes and graphene nanoribbons are intensely studied as
candidates for future electronic and optoelectronic devices.  In
particular, metallic tubes have some of the highest current densities
reported in any material~\cite{dekker00} and could lead to extremely
promising applications as electrical interconnects. However, in carbon
nanotubes or graphite, high currents
~\cite{lazzeri_PRL05,pop_PRL05,lazzeri_PRB06} or optical
excitations~\cite{kampfrath_PRL05,perfetti06} can induce a non-thermal
phonon distribution, with significant overpopulation of the optical
phonons E$_{2g}$ at ${\bm \Gamma}$ and A$'_1$ at {\bf K}. These {\it
hot phonons} develop because of a slower anharmonic decay rate with
respect to the generation rate~\cite{lazzeri_PRB06}, and cause a
significant reduction of the ballistic conductance of carbon nanotubes
at bias potentials larger than $\sim$ 0.2 V, severely limiting
interconnect
performance~\cite{dekker00,lazzeri_PRL05,pop_PRL05,lazzeri_PRB06}.  A
microscopic characterization of phonon decays~\cite{time_dom} is thus
a key step in improving the transport properties of these materials,
whereas engineering individual decay channels would allow to control
energy relaxation and ultimately performance.

Information on the phonon scattering mechanisms can be obtained from
Raman or infra-red (IR) measurements of the linewidths and line shifts
of the phonon modes~\cite{menendez84}. Indeed, the intrinsic linewidth
$\gamma^{in}$ in a defect-free sample is $\gamma^{in}=\gamma^{ep} +
\gamma^{pp}$, where $\gamma^{ep}$ and $\gamma^{pp}$ represent the
electron-phonon (EP) and anharmonic phonon-phonon (PP)
interactions~\cite{shukla93,debernardi95}.  The shift with temperature
of the harmonic phonon frequencies is also due to PP
interactions~\cite{cowley68,menendez84,shobhana91}.  While
experimental measurements are now available on graphene, graphite and
carbon nanotubes, their interpretation is not always straightforward.
For example, graphene has a E$_{2g}$ at ${\bm \Gamma}$ Raman-active
mode (the $G$ band) with a linewidth of $\sim$ 13
cm$^{-1}$~\cite{lw_graphene}.  In graphite this phonon splits in two
nearly-degenerate modes: the Raman-active E$_{2g}$ and the IR-active
E$_{1u}$ \cite{nemanich_SSC77}.  The linewidth of the Raman-active
mode (11.5~cm$^{-1}$ \cite{lw_graphite}) remains similar to that of
graphene, suggesting a negligible effect of the interactions among
different graphitic planes.  On the other hand, IR measurements show
that the linewidth of the E$_{1u}$ mode is much smaller
($\sim$4~cm$^{-1}$~\cite{nemanich_SSC77}).  The finite-temperature
line shift of E$_{2g}$ is also puzzling: recent experimental results
have shown very little difference between graphite and
graphene~\cite{ferrari}, while first-principles calculations find a
room-temperature in-plane thermal expansion coefficient for graphene
more than three times as large as that of graphite (both are
negative)~\cite{mounet05}.  Prompted by these results, and by the
central role played by phonon decays in controlling inelastic losses,
we characterize here the EP and PP scattering parameters of the
E$_{2g}$, E$_{1u}$ and A$'_1$ modes in graphite or graphene using
state-of-the-art first-principles calculations.  These parameters are
then used to compute the linewidths and line shifts of the Raman and IR
bands, and the PP decay lifetimes.

\begin{figure}
\centerline{\includegraphics[width=75mm]{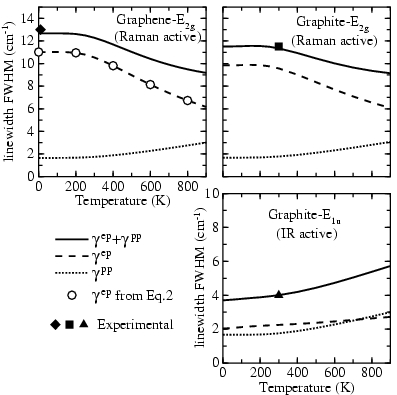}}
\caption{Total linewidths (solid line) for the E$_{2g}$ and E$_{1u}$
modes in graphene and graphite and their EP
($\gamma^{ep}$, dashed line) and PP ($\gamma^{pp}$, dotted
line) contributions, together with the results of Eq.~\ref{eq3}
(circles).  Measurements from
Refs.~\protect\cite{lw_graphene,lw_graphite,nemanich_SSC77} }.
\label{fig1}
\end{figure}

All the calculations are performed using density-functional theory
(DFT) and density-functional perturbation theory (DFPT)~\cite{dfpt} as
implemented in the {\sc PWSCF} package of the {\sc Quantum-ESPRESSO}
distribution~\cite{pwscf}. We use the local-density
approximation~\cite{lda}, norm-conserving
pseudopotentials~\cite{pseudi} and a plane-wave expansion up to a 55~Ry
cut-off.  Brillouin-zone sampling is performed on
32$\times$32$\times$8 and 32$\times$32$\times$1 Monkhorst-Pack meshes
for graphite and graphene, with a Fermi-Dirac smearing in the
electronic occupations of 0.02~Ry.  For graphite, the equilibrium
lattice parameters $a$ = 2.43~\AA~and $c/a$ = 2.725 are
used~\cite{mounet05}; for graphene, an interlayer spacing of 7~\AA~is
adopted.  The DFT accuracy in calculating vibrational properties in
graphite even in the presence of van der Waals interactions is
discussed in Ref.~\cite{mounet05}. The phonon frequencies, dynamical
matrices, and EP matrix elements are obtained using DFPT.  The phonon
anharmonic self-energy is given, at the lowest order in the
perturbative expansion with respect to the atomic mass, by the
tadpole, loop and bubble diagrams~\cite{lazzeri03} corresponding to
three- and four-phonon scattering terms.  Thus, we calculate the
third- and fourth-order derivatives of the total energy with respect
to atomic displacements; the former are obtained from DFPT~\cite{dfpt,lazzeri02}, while the
latter from finite differences over the relevant phonon eigenvectors.
The dynamical matrices are computed on a 16$\times$16$\times$1 or a
8$\times$8$\times$4 mesh (for graphene and graphite, respectively),
higher-order derivatives on 4$\times$4$\times$1 or 4$\times$4$\times$2
meshes.  Fourier interpolation~\cite{dfpt} then provides all these
quantities on finer grids (200$\times$200$\times$50 and
200$\times$200$\times$1 for graphite and graphene), over which we
perform all numerical integrations.

At the lowest order, a phonon acquires a finite linewidth by decaying
into two lower-energy phonons ($\gamma^{pp}$) or by creating an
electron-hole pair ($\gamma^{ep}$).  The PP contribution $\gamma^{pp}$
is given by the imaginary part of the phonon self-energy
$\Pi$~\cite{menendez84,lazzeri03}, which is determined by 3-phonon
scattering processes.  In the electron-hole creation process, a phonon
with wavevector $\bf q$ excites an electronic state $|{\bf k}i\rangle$
with wavevector ${\bf k}$ into the state $|({\bf k+q})j\rangle$.  The
scattering probability is thus given by the EP coupling
matrix element $g_{({ \bf k}+{\bf q})j,{\bf
k}i}$~\cite{piscanec04}.  According to the Fermi golden rule
\cite{allen}
\begin{eqnarray}
\gamma^{ep}_{{\bf q}}(T) &=& \frac{4\pi}{ N_{\rm k}}
\sum_{{\bf k},i,j} |g_{({\bf k}+{\bf q})j,{\bf k}i}|^2
[f_{{\bf k}i}(T)-f_{({\bf k}+{\bf q})j}(T)] \times \nonumber \\
&&\delta[\epsilon_{{\bf k}i} - \epsilon_{({\bf k}+{\bf q})j}
+ \hbar\omega_{ {\bf q}}],
\label{eq2}
\end{eqnarray}
where $\omega_{{\bf q}}$ is the phonon frequency, the sum is on
$N_{\rm k}$ ${\bf k}$ vectors, $f_{{\bf k}i}(T)$ is the Fermi-Dirac
occupation at temperature T for an electron with energy
$\epsilon_{{\bf k}i}$, and $\delta$ is the Dirac delta~\cite{note00}
(throughout the paper we will consider full-width at half-maximum
(FWHM) linewidths).

We report in Fig.~\ref{fig1} the linewidths for the E$_{2g}$ and
E$_{1u}$ modes in graphite and graphene, computed according to the
aforementioned procedure. Very good agreement is found with respect to
measurements~\cite{lw_graphene,lw_graphite,nemanich_SSC77}.  More
importantly, our calculations show that the phonon linewidth for the
E$_{2g}$ mode, and its dependence on temperature, is completely
dominated by the EP coupling, with a {\it decreasing linewidth as a
function of temperature}.  This effect is due to the strong $T$
dependence of $\gamma^{ep}$, which is only partially compensated by
$\gamma^{pp}$.

In order to rationalize this result, we consider a simplified model
for the temperature dependence of $\gamma^{ep}$ for the E$_{2g}$
modes: we assume a linearized band dispersion around the Fermi energy
($\epsilon_F$) and a model EP coupling~\cite{piscanec04}.
It can be easily shown (e.g. following Eq. 3 in Ref. \cite{lw_graphite})
that at finite $T$
\begin{equation}
\gamma^{ep}(T)=
\gamma^{ep}(0) 
\left[
f\left(-\frac{\hbar\omega_0}{2 k_BT}\right)-
f\left( \frac{\hbar\omega_0}{2 k_BT}\right)
\right],
\label{eq3}
\end{equation}
where, from DFT, $\gamma^{ep}(0)=11.01$~cm$^{-1}$ \cite{lw_graphite},
$\hbar\omega_0$=196~meV is the E$_{2g}$ phonon energy, $k_B$ is the
Boltzmann constant and $f(x)=1/[\exp(x)+1]$.  Eq.~\ref{eq3} reproduces
very well the full calculation of Eq.~\ref{eq2} (see Fig.~\ref{fig1})
and can be used to understand the temperature dependence of
$\gamma^{ep}(T)$, since this is now proportional to the difference
between the occupations of states below and above $\epsilon_F$.  As
$T$ increases, the occupation of filled states below $\epsilon_F$
decreases, while the empty states are occupied more, resulting in the
observed decrease of $\gamma^{ep}(T)$ with temperature.

\begin{figure}
\centerline{\includegraphics[width=85mm]{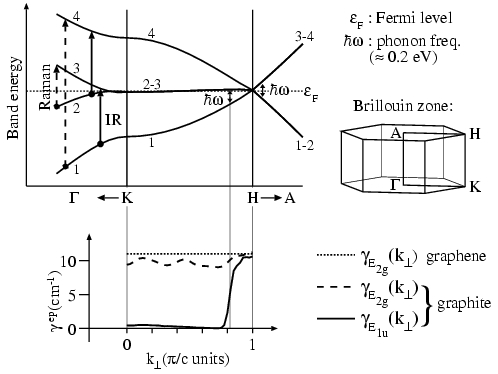}}
\caption{EP coupling contributions to the linewidths
($\gamma^{ep}$) of the E$_{2g}$ (Raman active) and E$_{1u}$ (IR
active) ${\bm \Gamma}$ modes.  Upper panel: the four $\pi$ bands of
graphite.  Lower panel: Decomposition of $\gamma^{ep}$ into the
different $k_{\perp}$ contributions.}
\label{fig2}
\end{figure}

It is important to note that $\gamma^{ep}(0)$ for the E$_{1u}$ mode in
graphite is almost five times smaller than for the case of
E$_{2g}$. This difference can be understood by decomposing
Eq.~\ref{eq2} in parallel and perpendicular contributions, where ${\bf
k}_{\perp}$ is the component perpendicular to the graphene planes and
${\bf k_{\parallel}}$ is the in-plane projection.  We define
$\gamma(k_{\perp})$ as the contribution to the EP linewidth obtained
from Eq.~\ref{eq2} when restricting the k-point integration on those
vectors ${\bf k}$ that satisfy $\hat{\bf c}\cdot {\bf k}={
k}_{\perp}$, where $\hat{\bf c}$ is the unit vector perpendicular to the
graphene planes.  With such definition $\gamma^{ep}=\int_0^1
\gamma(k_{\perp})~ dk_{\perp}$, where $k_{\perp}$ is in units of
$\pi/c$.  The electronic states with a non-zero contribution to
Eq.~\ref{eq2} are those allowed by energy conservation and by a
non-zero EP coupling.  Energy conservation alone selects the four
$\pi$ bands near the Fermi level (labeled 1 to 4, from the lowest to
the highest, in Fig.~\ref{fig2}).  For the E$_{1u}$ mode the computed
EP coupling allows mainly transitions from band 1 to 3 and from band 2 to
4. Since the minimum gap between bands 1 and 3 (and 2 and 4) varies
considerably as a function of $k_{\perp}$, and the IR transition
satisfies energy conservation only for $k_{\perp}\gtrsim 0.8$
(Fig.~\ref{fig2}), we have a small $\gamma^{ep}$ for this mode.  On
the contrary, for the E$_{2g}$ mode the EP coupling allows mainly transitions
from band 1 to 4 and from band 2 to 3; energy-conservation means that
only the transition between 2 and 3 is essentially active.  Since the
minimum gap between bands 2 and 3 is always zero, the transition is
active for any $k_{\perp}$. It turns out that $\gamma_{{\rm
E}_{2g}}(k_{\perp})$ is almost a constant and its value is similar to
that of graphene (Fig.~\ref{fig2}) and much larger than that for the
IR-active mode ($\gamma^{ep}_{{\rm E}_{2g}}\sim5\gamma^{ep}_{{\rm
E}_{1u}}$). Interstingly, the IR-active mode of a graphene bilayer
should have a vanishing $\gamma^{ep}$, since the bilayer bands are
very similar to those of graphite with $k_{\perp}=0$.

\begin{figure}
\centerline{\includegraphics[width=85mm]{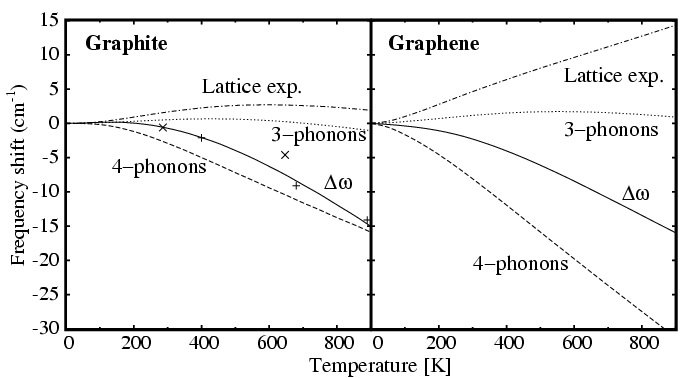}}
\caption{Line shifts for the E$_{2g}$ ${\bm \Gamma}$ phonons in
graphene and graphite: total shift, $\Delta\omega$, 
solid lines (see footnote~\cite{fit_ls} for a polynomial fit to
$\Delta\omega$); 3-phonon, 4-phonon
and thermal expansion contributions, dotted, dashed and
dot-dashed lines, respectively. At 0 K, the 3-phonon and 4-phonon contributions
for graphite (graphene) are -14.06 cm$^{-1}$ and 1.30
cm$^{-1}$ (-14.02 cm$^{-1}$ and 3.03 cm$^{-1}$), respectively.
Experimental data for graphite ($+$, Ref.~\cite{kagi_GEO94} and $\times$,
Ref.~\cite{tan_APL99}) are also shown.}
\label{fig3}
\end{figure}

The temperature-dependent line shift is another quantity that is
easily accessible by e.g. Raman spectroscopy, and that
provides powerful information on the anharmonicity.  The PP
contribution to line shifts is given by the real part of the
self-energy $\Pi$~\cite{lazzeri03}; as mentioned before, at the lowest
order this includes both 3-phonon and 4-phonon scattering terms.  A
further contribution descends straightforwardly from the lattice
thermal expansion, that is especially large and negative in graphene
\cite{mounet05}.  This contribution is obtained by computing the
harmonic frequency at the lattice parameter appropriate to the given
temperature, obtained within the quasi-harmonic
approximation~\cite{mounet05}.
Within the present approach the EP contribution to the frequency shift
is taken into account exactly (within DFT) in the harmonic
frequencies~\cite{dfpt}.


Fig.~\ref{fig3} shows our computed line shifts for the E$_{2g}$
mode~\cite{eph_T}.
The results are in good agreement with available experiments
\cite{kagi_GEO94,tan_APL99}, and in excellent agreement with recent
measurements for graphite and graphene \cite{ferrari}.  In both cases
the frequencies shift down with temperature---an unusual result for an
optical mode where the bond-bond distances are predicted to become
shorter with temperature. In reality, lattice contraction does provide
the expected upward shift---and a much larger one for graphene than
for graphite.  Still, the overall behavior is dominated by a downshift
driven by the 4-phonon scattering term, almost two times stronger in
graphene than in graphite (Fig.~\ref{fig3}).  Thus, while the
individual anharmonic contributions in graphite and graphene are quite
different, the E$_{2g}$ line shifts are always downwards (driven by the
4-phonon contributions) and very similar in the two systems thanks to
the compensation between different but opposite contributions.

\begin{figure}
\centerline{\includegraphics[width=85mm]{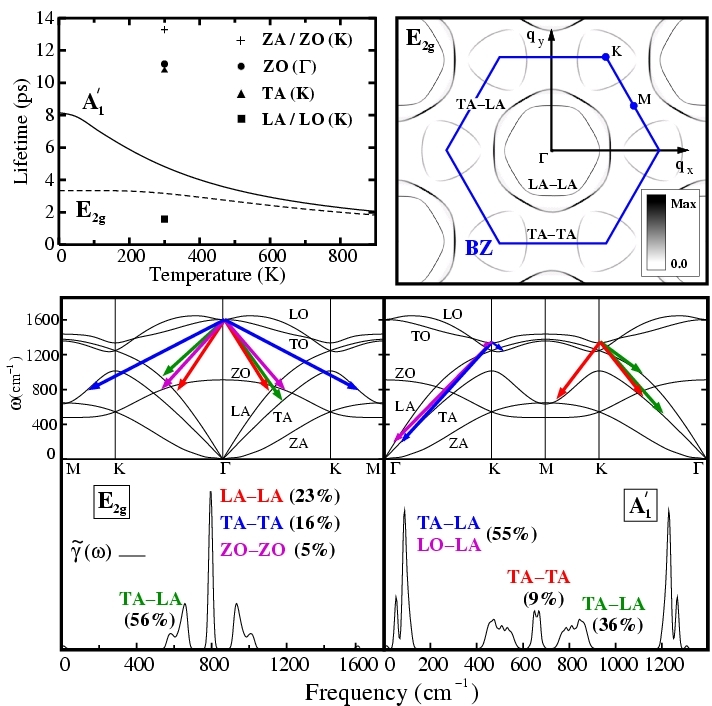}}
\caption{(Color online) Upper left panel: computed PP lifetime of the
E$_{2g}$ and A$'_1$ modes of graphene. We also report the computed
lifetimes (300 K) of the other modes at ${\bm \Gamma}$ and {\bf K}.
Upper right panel: probability per unit time (300 K) that the E$_{2g}$
mode at ${\bm \Gamma}$ decays into two phonons of wavevectors {\bf q}
and $-${\bf q}. Upper part of the lower panel: schematic
representation of the different PP decay channels for the modes
E$_{2g}$ at ${\bm \Gamma}$ (left) and A$'_1$ at {\bf K}
(right). Below, probability per unit time $\tilde{\gamma}$ that the
E$_{2g}$ or the A$'_1$ modes (of frequency $\omega_0$) decay into two
modes of frequencies $\omega$ and $\omega_0$-$\omega$. The relative
weight of the decay channels is also reported.  }
\label{fig4}
\end{figure}

Finally, we focus on the analysis of the anharmonic phonon decay
processes.  We show in Fig.~\ref{fig4} the anharmonic phonon lifetimes
($\tau=1/\gamma^{pp}$) and the decay channels for the modes E$_{2g}$
at ${\bm \Gamma}$ and A$'_1$ at {\bf K} in graphene---these are the
two modes with the strongest EP coupling, and the ones that will be
overpopulated during steady-state operation in an interconnect
\cite{lazzeri_PRB06} (the results for graphite are very similar, see
Fig.~\ref{fig1}).  The values obtained are of the same order of the
optical-phonon thermalization time (7~ps) estimated in graphite from
time-resolved terahertz spectroscopy ~\cite{kampfrath_PRL05}.  This
result is also in agreement with the empirical choice of
Ref.~\cite{lazzeri_PRB06} where the experimental I-V characteristic of
metallic-tube interconnects was modeled by a coupled Boltzmann
transport equation for phonons and electrons, assuming $\tau_{\bm
\Gamma}=\tau_{\bf K}\sim 5$~ps.  The values of $\tau_{\bm \Gamma}$ and
$\tau_{\bf K}$ that we obtain from first-principles confirm this
assumption, but provide much needed novel insight on the relative
relevance of the different decay processes.  In particular, it is
found that $\tau_{\bf K}>\tau_{\bm \Gamma}$ (Fig.~\ref{fig4}); in
addition, since the EP coupling for the {\bf K} mode is stronger than for
${\bm \Gamma}$ \cite{lazzeri_PRB06}, we find that the phonon
population at {\bf K} will be dominant in determining inelastic
losses, with the high-bias resistivity due to scattering with {\bf K}
phonons.  Moreover, $\tau_{\bf K}$ has a large decay channel towards
low-energy acoustic phonon modes (Fig.~\ref{fig4}, bottom right panel)
that is not available to ${\bm \Gamma}$ phonons.  This means that a
strong temperature-dependence is present in the typical operation
range of 100-500~K, and that the population of acoustic phonons can
strongly affect hot phonons and transport properties. Since acoustic
phonons have a lower thermal-impedance mismatch with the substrate, it
is expected that efficient thermalization strategies should focus on
engineering the optimal coupling with the substrate. The
present results and inclusion of acoustic phonons in the model of
Ref.~\cite{lazzeri_PRB06} should provide a realistic
ab-initio description of the coupled electronic and thermal dynamics
in carbon nanostructures \cite{bonini2020}.

%

In conclusion, we have presented a detailed analysis of anharmonic
effects in graphene and graphite, based on the explicit calculation
of the PP and EP interactions within
DFPT.  Excellent agreement with experimental
results---where available---is found.  We have explained the large
differences in the linewidths for the closely related Raman and IR
G-bands in graphite, and the closely similar line shifts for the G-band
in graphene and graphite, notwithstanding very different
thermal-expansion parameters.  Moreover, the anomalous decrease of the
Raman G-band linewidth with temperature, predicted for both graphene
and graphite, is rationalized through its dominant EP
contribution; the negative dependence on temperature is accurately
captured by a simple phenomenological expression.  The PP
decay channels for the critical vibrational excitations that limit
ballistic transport have been identified, with fundamental
consequences in understanding and engineering electronic transport in
metallic nanotubes and graphene ribbons interconnects.

We acknowledge support from the IFC Focus Center, a
Semiconductor Research Corporation program, and MITRE (N. B.),
and NSF DMR-0304019 (N. M.). Computational support was
provided by IDRIS (Orsay, project n$^o$ 071202), and NSF DMR-0414849
and PNNL EMSL-UP-9597.


\begin{thebibliography}{99}

\bibitem{dekker00}
Z. Yao, C.L. Kane, and C. Dekker,
Phys. Rev. Lett. {\bf 84}, 2941 (2000).

\bibitem{lazzeri_PRL05}
M. Lazzeri {\it et al.}, 
Phys. Rev. Lett. {\bf 95}, 236802 (2005).

\bibitem{pop_PRL05}
E. Pop {\it et al.},
Phys. Rev. Lett. {\bf 95}, 155505 (2005).

\bibitem{lazzeri_PRB06}
M.~Lazzeri and F.~Mauri,
Phys. Rev. B {\bf 73}, 165419 (2006).

\bibitem{kampfrath_PRL05}
T. Kampfrath {\it et al.}, Phys. Rev. Lett. {\bf 95}, 187403 (2005).

\bibitem{perfetti06}
L. Perfetti {\it et al.}
Phys. Rev. Lett. {\bf 96}, 027401 (2006).

\bibitem{time_dom}
B. F. Habenicht {\it et al.}, Phys. Rev. Lett. {\bf 96}, 187401 (2006);
Y. Miyamoto {\it et al.}, {\it ibid.} {\bf 97}, 126104 (2006);
A. Gambetta {\it et al.}, Nature Phys. {\bf 2}, 515 (2006).

\bibitem{menendez84}
J. Menendez and M. Cardona,
Phys. Rev. B {\bf 29}, 2051 (1984).

\bibitem{shukla93}
A. Shukla {\it et al.},
Phys. Rev. Lett. {\bf 90}, 095506 (2003).

\bibitem{debernardi95}
A. Debernardi {\it et al.}, Phys. Rev. Lett. {\bf 75}, 1819 (1995).

\bibitem{cowley68} R. A. Cowley, Rep. Prog. Phys. {\bf 31}, 128 (1968).

\bibitem{shobhana91} S. Narasimhan and D. Vanderbilt,
Phys. Rev. B {\bf 43}, 4541 (1991).

\bibitem{lw_graphene}
J. Yan {\it et al.}, Phys. Rev. Lett. {\bf 98}, 166802 (2007).
The spectral resolution of 2~cm$^{-1}$ has been subtracted
from the bare measurements.

\bibitem{nemanich_SSC77}
R.J. Nemanich {\it et al.}, Sol. Stat. Comm. {\bf 23}, 117 (1977).

\bibitem{lw_graphite}
M. Lazzeri {\it et al.},
Phys. Rev. B {\bf 73}, 155426 (2006).

\bibitem{ferrari}
A. C. Ferrari, private communication; I. Calizo {\it et al.}, arXiv:0708.1223v1.



\bibitem{mounet05}
N. Mounet and N. Marzari,
Phys. Rev. B {\bf 71}, 205214 (2005).

\bibitem{dfpt}
S. Baroni {\it et al.}, Rev.\ Mod.\ Phys.\ {\bf 73}, 515 (2001).

\bibitem{pwscf}
S. Baroni {\it et al.}, http://www.quantum-espresso.org

\bibitem{lda}
D.M. Ceperley and B.J. Alder, Phys. Rev. Lett. {\bf 45}, 566 (1980).

\bibitem{pseudi}
N. Troullier and J. L. Martins, Phys. Rev. B 43, 1993 (1991);
M. Fuchs and M. Scheffler, Comput. Phys. Commun. {\bf 119}, 67 (1999).

\bibitem{lazzeri03}
M. Lazzeri, M. Calandra, and F. Mauri,
Phys. Rev. B {\bf 68}, 220509(R) (2003).

\bibitem{lazzeri02}
M. Lazzeri and S. de~Gironcoli, Phys. Rev. B {\bf 65}, 245402 (2002).

\bibitem{piscanec04}
S. Piscanec {\it et al.},
Phys. Rev. Lett. {\bf 93}, 185503 (2004).

\bibitem{allen}
P.B. Allen, Phys. Rev. B {\bf 6}, 2577 (1972);
P.B. Allen, R. Silberglitt, {\it ibid.} {\bf 9}, 4733 (1974).

\bibitem{note00}
The {\bf k}-point sum in Eq.~\protect\ref{eq2} is performed on a
2430$\times$2430$\times$144 grid for graphite (and
2430$\times$2430$\times$1 for graphene), using for $\delta$ a
Lorentzian with a 0.13 mRy (20~K) width.  To reduce the computational
costs, this sum can be
restricted to a small region around the {\bf K-H} line where the
addends are different from zero.


\bibitem{eph_T}
The EP contribution changes by less than 1~cm$^{-1}$ in the 0-800
K temperature range.

\bibitem{kagi_GEO94}
H. Kagi {\it et al.},
Geochim. Cosmochim. Acta {\bf 58}, 3527 (1994).

\bibitem{tan_APL99}
P. H. Tan {\it et al.},
Appl. Phys. Lett. {\bf 74}, 1818 (1999).

\bibitem{bonini2020} N. Bonini {\it et al.} in preparation.

\bibitem{fit_ls}
A polynomial of the form $aT^2 + bT^3 + cT^4 + dT^5$
fits well (especially between 150-900K) the {\it ab initio} results,
with $a=2.2679\times10^{-5}$, $b=-1.4836\times10^{-7}$,
$c=1.7869\times10^{-10}$ and $d=-7.1998\times10^{-14}$ for graphite
and $a=-2.6595\times10^{-5}$, $b=-1.3568\times10^{-9}$,
$c=1.3633\times10^{-11}$ and $d=-4.0258\times10^{-15}$ for graphene.

\end{thebibliography}
\end{document}